# The effect of the rapid growth of covid-19 publications on citation indicators

Peter Sjögårde*

* *peter.sjogarde@ki.se*
Health Informatics Centre, Department of Learning, Informatics, Management and Ethics
Karolinska Institutet
University Library, Karolinska Institutet, Stockholm, Sweden
https://orcid.org/0000-0003-4442-1360

**Abstract**
A concern has been raised that "covidization" of research would cause an overemphasizes on covid-19 and pandemics at the expense of other research. The rapid growth of publications related to the Covid-19 pandemic renders a vast amount of citations from this literature. This growth may affect bibliometric indicators. In this paper I explored how the growth of covid-19 publications influences bibliometric indicators commonly used in university rankings, research evaluation and research allocation, namely the field normalized citation score and the journal impact factor.

I found that the burst of publications in the early stage of the covid-19 pandemic affects field-normalized citation scores and will affect the journal impact factor. Publications unrelated to covid-19 are also heavily affected. I conclude that there is a considerable risk to draw misleading conclusions from citation indicators spanning over the beginning of the covid-19 pandemic, in particular when time series are used and when the biomedical literature is assessed.

**Introduction**
The pandemic caused by the sars-cov-2 virus, which causes the covid-19 infection, started in late 2019 and was followed by a rapid emergence and increase of publications addressing different aspects of the pandemic (Aristovnik et al., 2020; Colavizza et al., 2021). The rapid growth of publications in this topic also renders a vast amount of citations from this literature. A recent preprint shows that the most cited publications in 2020-2021 are heavily dominated by covid-19 publications and that the same relation is found when looking at the most cited researchers (Ioannidis et al., 2022). A concern has been raised that "covidization" of research would cause an overemphasizes on covid-19 and pandemics at the expense of other research, e.g. other global public health threats (Adam, 2020).

Bibliometric indicators are used in university rankings, resource allocation and research evaluations at various levels and thereby influence research policy. The rapid growth of covid-19 publications may affect bibliometric indicators. Sjögårde and Didegah (2022) has shown that growing research topics in general have a citation advantage over more slowly growing or declining research topics. This relation can partly be attributed to a simple quantitative aspect, namely that in a growing topic there are more citing publications than publications to be cited (Vinkler, 1996). Publications published shortly after the emergence of a topic therefore have a citation advantage.

In this paper I explore how the rapid growth of covid-19 publications influences bibliometric indicators commonly used in university rankings, research evaluation and research allocation, namely field normalized citation score and the journal impact factor (JIF).

**Characteristics of covid-19 publications**

The analyses were restricted to biomedical literature. A search query (Table 1) was used to identify publications in PubMed/MEDLINE about covid-19 or the virus causing the disease (sars-cov-2). The query has been elaborated by the university library at Karolinska Institutet and was conducted 26 January 2022. I refer to the set of publications retrieved by this search query as covid-19 publications.

Table 1: Search query for Covid-19/SARS-CoV2 research

| |
|---|
| COVID*[tw] OR nCov[tw] OR 2019 ncov[tw] OR novel coronavirus[tw] OR novel corona virus[tw] OR "COVID-19"[All Fields] OR "COVID-2019"[All Fields] OR "severe acute respiratory syndrome coronavirus 2"[Supplementary Concept] OR "severe acute respiratory syndrome coronavirus 2"[All Fields] OR "2019-nCoV"[All Fields] OR "SARS-CoV-2"[All Fields] OR "2019nCoV"[All Fields] OR (("Wuhan"[All Fields] AND ("coronavirus"[MeSH Terms] OR "corona virus"[All Fields] OR "coronavirus"[All Fields])) AND (2019/12[PDAT] OR 2020:2022[PDAT])) |

Citations were calculated using in-house version of the Web of Science (WoS) at Karolinska Institutet, containing data from the Science Citation Index Expanded, Social Sciences Citation Index and Arts & Humanities Citation Index.[1] The analyses are restricted to the intersection between Web of Science and PubMed if not otherwise stated. Furthermore, a restriction was made to the Web of Science publication types "article" and "review" and to publications from 2020 and 2021. The search resulted in 31,789 publications in 2020 and 74,064 in 2021, an increase of about 133%.

shows the number of publications in 2021 at the primary y-axis for the covid-19 set as well as the 20 largest Web of Science journal categories in medicine (without restriction to PubMed publications). The secondary axis shows the number of citations from 2021 to 2020. It is shown that covid-19 publications cite more publications from 2020 than any of the journal categories, despite being less than half the size of the largest category ("Materials Science, Multidisciplinary"). This shows that covid-19 publications cite recent publications in a much higher extent than other medical publications.

---

[1] Certain data included herein are derived from the Web of Science ® prepared by Clarivate Analytics ®, Inc. (Clarivate®), Philadelphia, Pennsylvania, USA: © Copyright Clarivate Analytics Group ® 2022. All rights reserved.

*Figure 1: Number of publications in 2021 (primary axis) and number of outgoing citations from 2021 to 2020 (secondary axis) for the covid-19 set and for the 20 largest Web of Science journal categories in medicine. Web of Science journal categories ordered by number of publications from left to right.*

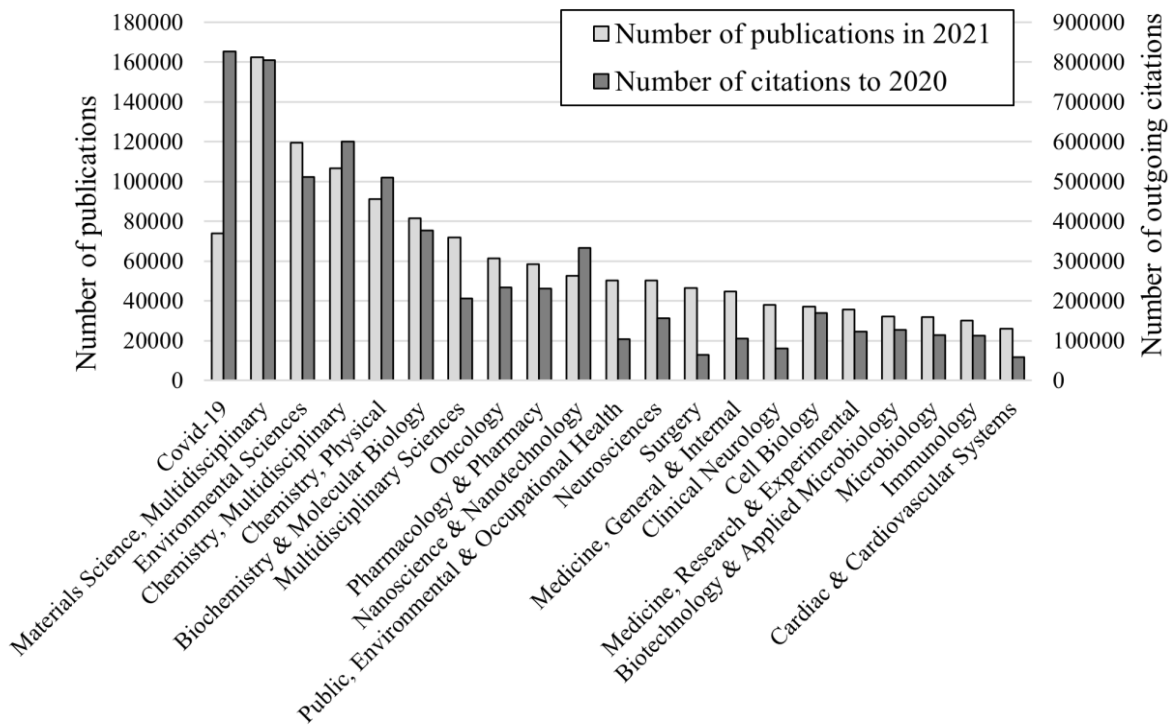

The average number of outgoing citations from covid-19 publications in 2021 is relativly low, about 30 (Figure 2). However, as much as 37% of these citations goes to the previous year. This figure varies between 5-15% in medical journal categories. It is clear from figure 1 and 2 that covid-19 publications cited recent papers to a much higher extent than expected in the beginning of the pandemic.

*Figure 2: Total number of outgoing citations in 2021 as well as share of citations to 2020 for the covid-19 set and for the 20 largest Web of Science journal categories in medicine. Web of Science journal categories ordered by number of outgoing citations to 2020 from left to right.*

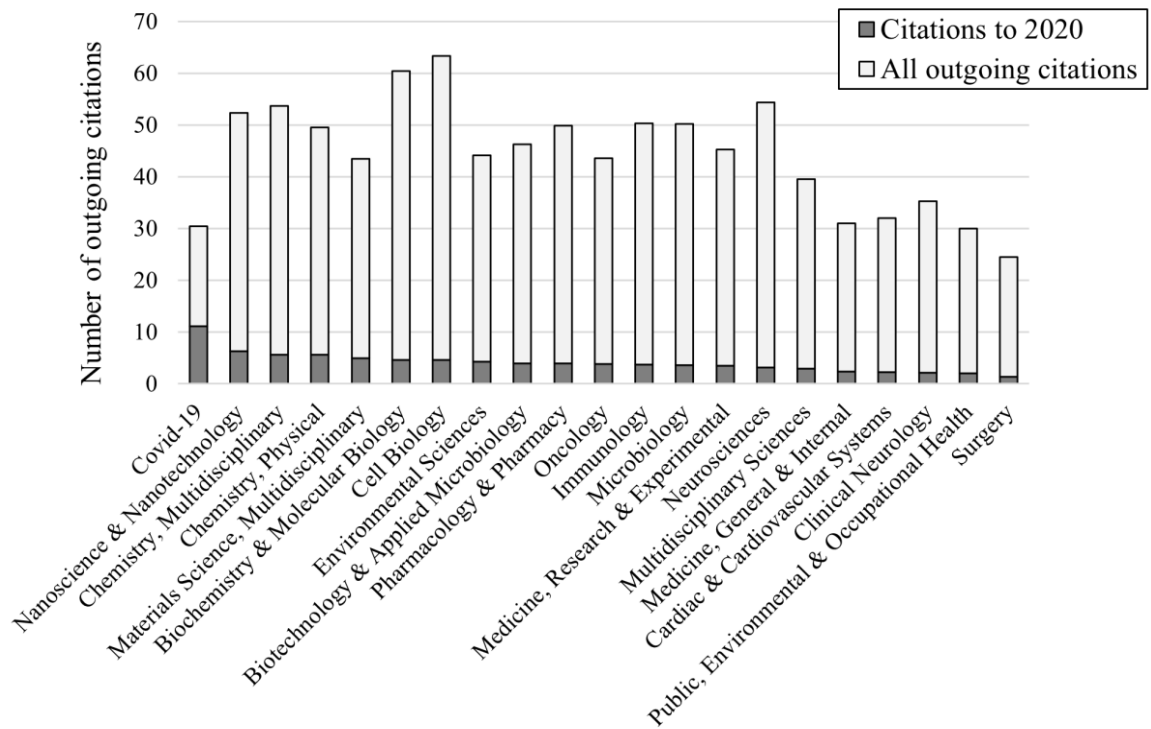

**Highly cited publications**

To be able to compare highly cited publications in 2020 with previous publication years a two-year citation window was used. Figure 3 shows publications from 2019 with more than 200 citations, ordered from left to right by the number of received citations. About 170 publications from 2019 had been cited more than 200 times before the end of 2020. Figure 4 shows the same information for publications from 2020 but with covid-19 publications excluded. This figure looks approximately the same as the figure for 2019. A small increase from 2019 and 2020 of the number of publications with more than 200 citations after a two-year period can be seen, when covid-19 publications are excluded. In Figure 5 the covid-19 publications have been added (yellow marks). The number of publications with more than 200 citations has now increased to almost 1,000. This shows the domination of covid-19 publications among the world's most highly cited publications.

*Figure 3: Publications from 2019 with more than 200 citations using a two-year time window. Order by rank from left to right.*

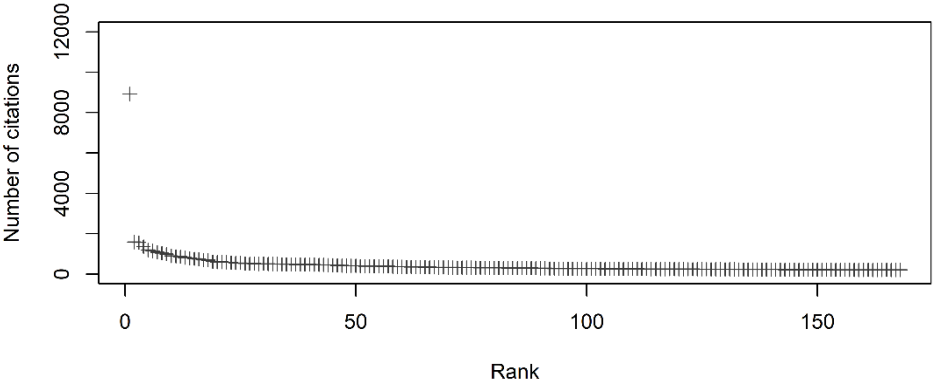

*Figure 4: Publications from 2020 with more than 200 citations using a two-year time window. Order by rank from left to right. Covid-19 publications excluded.*

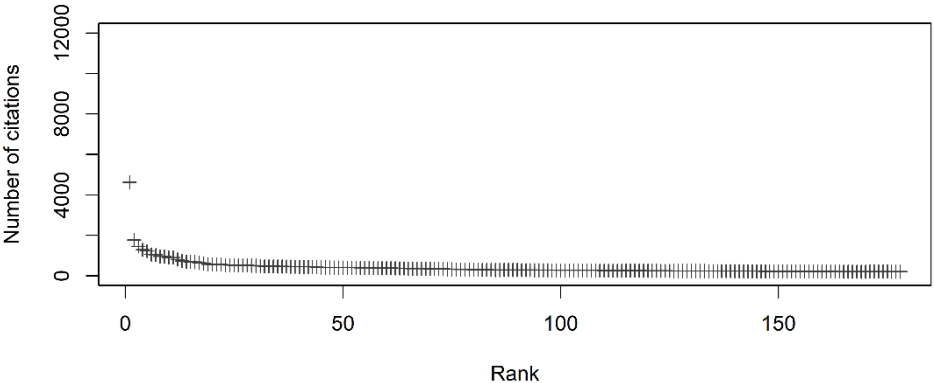

*Figure 5: Publications from 2020 with more than 200 citations using a two-year time window. Order by rank from left to right. Covid-19 publications in yellow.*

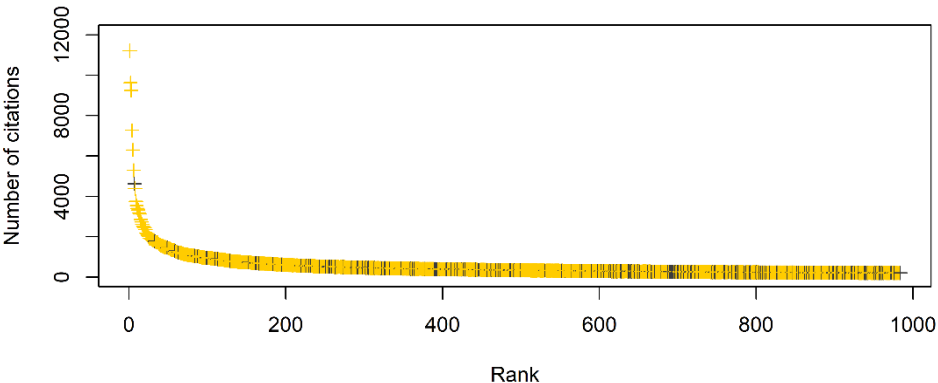

**Effects on field normalization**

The citation rate of covid-19 publications in 2020 was in general much higher than other publications within the same journal category after a two-year period (Figure 6). For example, in "Medicine, General & Internal" the average citation rate was about 3.5 for non-covid publications, 55 for covid-19 publications and 7,7 for the category in total.

The field normalized citation score is used in many rankings and evaluations. For a publication, this indicator is usually calculated by dividing the number of citations to the publication by the average number of citations in the same field, from the same publication year and commonly also of the same publication type. The Web of science journal categories are commonly used as fields in this calculation. If we again take "Medicine, General & Internal" as an example, we see that the value of the denominator in this field (the field reference value) is more than doubled by the covid-19 publications. To illustrate the effect on the field normalized citation score, we can for example consider a cancer publication having no relation to covid-19 and which has received 14 citations after a two-year period. This publication would have a field normalized citation score of 4 if covid-publications are excluded. However, the covid-19 publications reduce the normalized citation rate of such publication to about 1.8.

*Figure 6: Average number of citations in 2020 for the covid-19 set and for the 20 largest Web of Science categories in medicine after a two-year period. For each category the dark grey bar shows the average number of citations for all publications, the yellow bar shows the average number of citations for covid-19 publications and the light grey bar shows the average number of citations for non-covid publications. Web of Science categories are ordered from left to right by the proportion of covid-19 publications.*

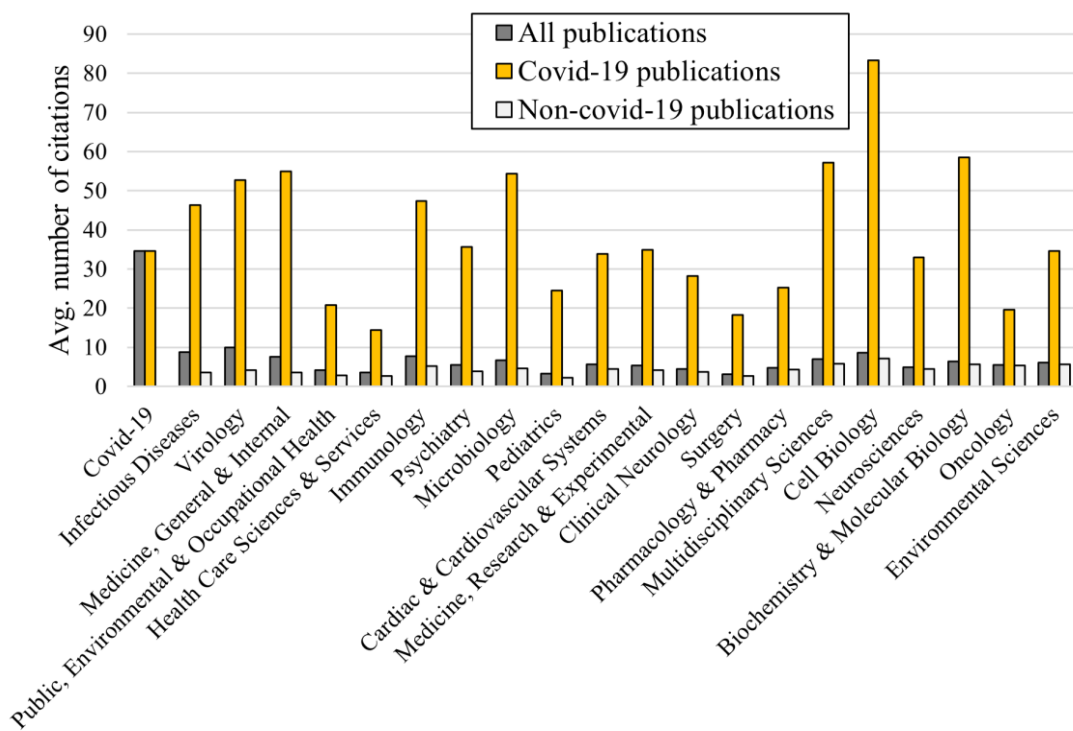

Even in some fields with a low proportion of covid-19 publications the average citation rate is heavily affected by the high citation rates of covid-19 publications. In for example "Cardiac &

Cardiovascular Systems" with only about 3.9% covid-19 publications (Table 1) the average citation rate is about 24% higher when covid-19 publications are considered (5.6 compared to 4.5).

*Table 2: The total number of publications, covid-19 publications and non-covid-19 publications in 2020 for the covid-19 set and for the 20 largest Web of Science categories in medicine.*

| Category | Publications | Non-covid-19 publications | Covid-19 publications |
|---|---|---|---|
| Covid-19 | 31789 | 0 | 31789 (100%) |
| Infectious Diseases | 19174 | 16807 | 2367 (12.3%) |
| Virology | 6897 | 6063 | 834 (12.1%) |
| Medicine, General & Internal | 40691 | 37414 | 3277 (8.1%) |
| Public, Environmental & Occupational Health | 46981 | 43492 | 3489 (7.4%) |
| Health Care Sciences & Services | 16678 | 15573 | 1105 (6.6%) |
| Immunology | 29510 | 27713 | 1797 (6.1%) |
| Psychiatry | 24584 | 23382 | 1202 (4.9%) |
| Microbiology | 28791 | 27581 | 1210 (4.2%) |
| Pediatrics | 19955 | 19120 | 835 (4.2%) |
| Cardiac & Cardiovascular Systems | 23421 | 22501 | 920 (3.9%) |
| Medicine, Research & Experimental | 37055 | 35679 | 1376 (3.7%) |
| Clinical Neurology | 35118 | 34119 | 999 (2.8%) |
| Surgery | 41290 | 40119 | 1171 (2.8%) |
| Pharmacology & Pharmacy | 53386 | 51940 | 1446 (2.7%) |
| Multidisciplinary Sciences | 73644 | 71916 | 1728 (2.3%) |
| Cell Biology | 36271 | 35553 | 718 (2%) |
| Neurosciences | 44084 | 43424 | 660 (1.5%) |
| Biochemistry & Molecular Biology | 74850 | 73774 | 1076 (1.4%) |
| Oncology | 55057 | 54291 | 766 (1.4%) |
| Environmental Sciences | 100689 | 99531 | 1158 (1.2%) |

**Effects on the journal impact factor (JIF)**

JIF is used in many rankings and evaluations, despite such use has been discussed and criticized (e.g., see Aksnes et al., 2019; Hicks et al., 2015; Seglen, 1998). JIF is "calculated by dividing the number of current year citations to the source items published in that journal during the previous two years" ("The Clarivate Analytics Impact Factor," n.d.). The most recent impact factor is the 2021 edition, which is based on citations from 2020 to 2018-2019. This edition has not been much affected by the covid-19 pandemic, because there were practically no covid-19 publications in 2018-2019 and only about 32,000 covid-19 publications in 2020. However, the burst of covid-19 publications in 2021 citing publications from 2020 is likely to affect the 2022 edition of the indicator, which will be based on citations from 2021 to publications from 2019-2020.

I did not have access to all the indexes in Web of Science that are used to calculate JIF. To calculate an estimated JIF, I used the three main indexes (Science Citation Index Expanded, Social Sciences Citation Index and Arts & Humanities Citation Index) across 8,037 journals (journals with at least 50 articles or reviews). The estimated JIF for year y was calculated as the total number of citations (no restrictions to publication type) from year y to the journal, divided by the number of publications of the types "article" and "review" in year y-1 and y-2. Since the estimated JIF is based on fewer citing publications it is generally lower than the real JIF. However, this does not have much effect on the relative ranking of journals. Figure 7 shows the correlation between the real JIF in 2020 (edition 2021) and the estimated JIF for the same year.

*Figure 7: Correlation between estimated JIF and real JIF based on citations from 2020 to 2018-2019 (edition 2021).*

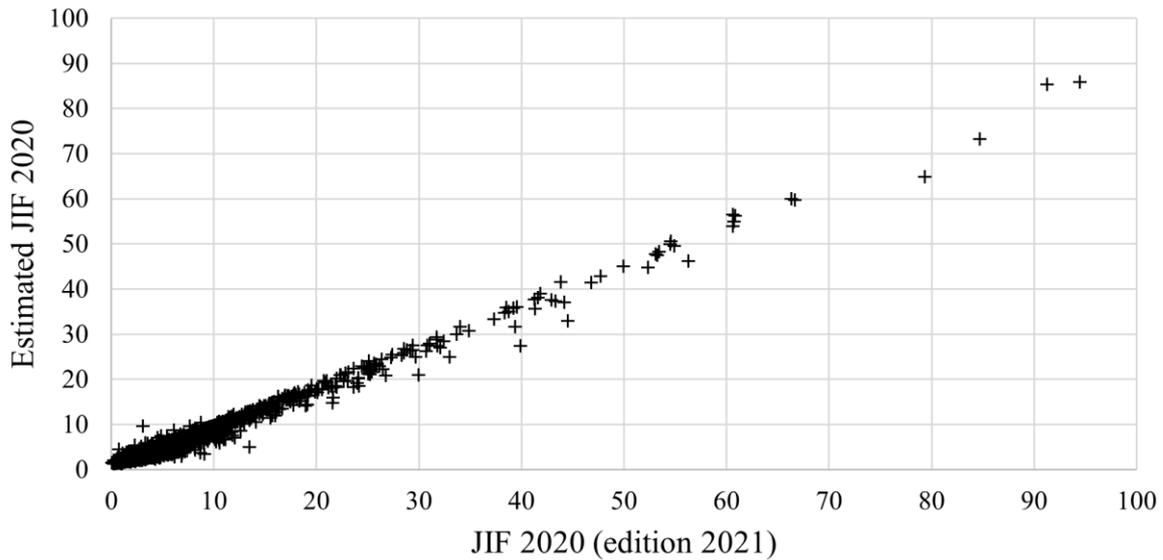

Most journals with a high proportion of covid-19 publications increased their estimated JIF between 2020 and 2021. Figure 8 shows the about 8,000 journals in a scatter plot with the percentual change of the JIF between 2020 and 2021 on the y-axis and the percentage of covid-19 publications on the x-axis. Almost all the journals with more than 10% covid-19 publications increased their JIF. 18 journals with more than 5% covid-19 publications (out of 521) increased their JIF by more than 200%. Only three journals with less than 5% covid-19 publications (out of 7,516) increased their JIF by as much.

*Figure 8: Correlation between the percentage of covid-19 publications in 2019-2020 (x-axis) and increase of estimated JIF between 2020 and 2021 (y-axis) for 8,037 journals in Web of Science.*

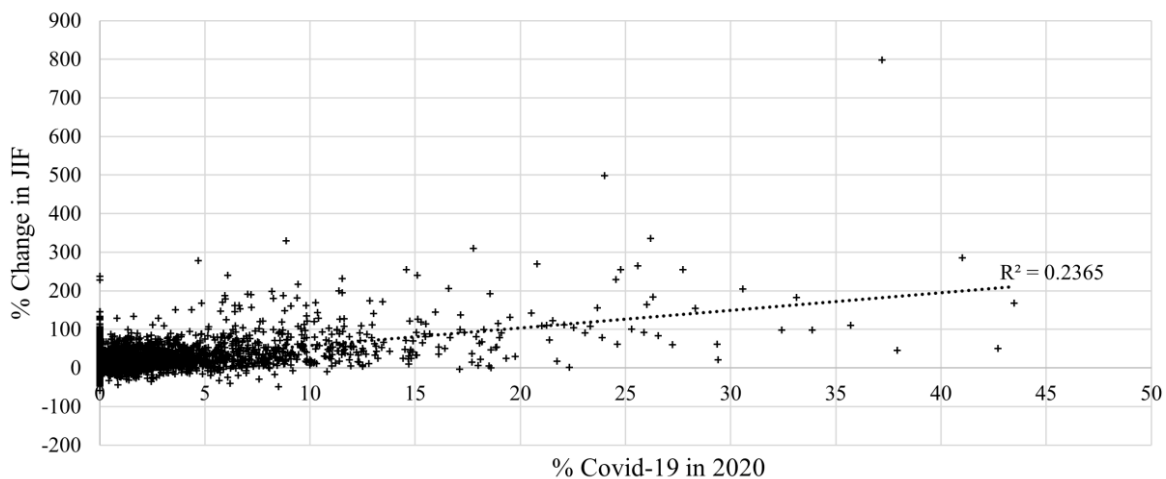

Most of the 20 journals having the highest absolute increase of the estimated JIF had between 5 and 15% covid-19 publications in 2019-2020. "Lancet" goes from an estimated JIF of about 65 to 169 and is predicted to be the journal with the highest JIF in the 2022 edition.

*Figure 9: The 20 journals having the highest absolute increase of the estimated JIF between 2020 (based on citations from 2020 to 2018-2019) and 2021 (based on citations from 2021 to 2019-2020). The journal name is followed by the share of covid-19 publications in 2019-2020.*

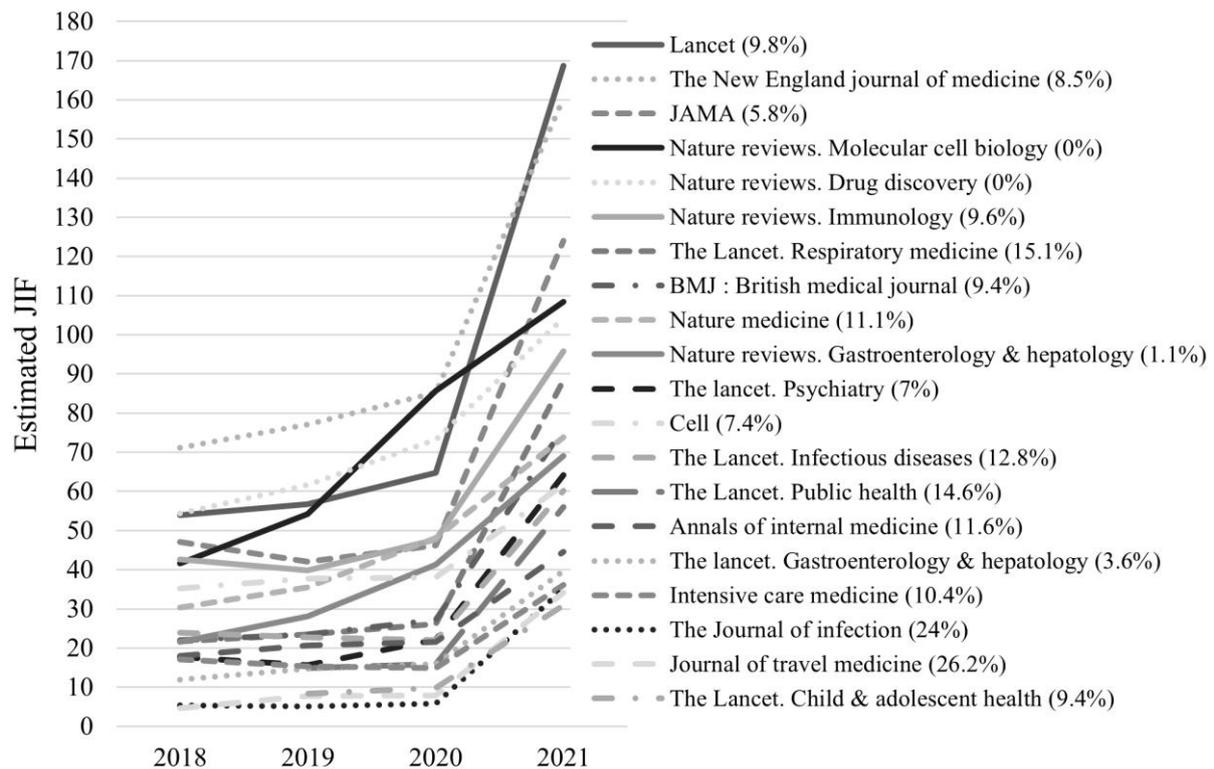

**Discussion and conclusions**

The burst of publications in the early stage of the covid-19 pandemic addressing topics related to the pandemic has profound consequences on citation indicators, severely affecting the validity of bibliometric indicators. The study has shown that field normalized citation scores may be distorted, and the JIF was estimated to be profoundly affected.

A hypothetical unit can be used to illustrate how units under assessment may be affected in evaluations. Consider a unit with a constant publishing output completely in the field of "Pediatrics" and with an average citation rate corresponding to the field average. Such unit would have a field normalized citation score of 1, given that full counting has been applied. Now assume that this unit keeps publishing non-covid publications in 2020 with the same average citation rate as the earlier publications. Its field normalized citation score will now drop from 1 to 0.71 because of the change of the field reference value caused by the covid-19 publications. Now consider that this unit publishes equal parts in the three largest journals for pediatrics ("JAMA pediatrics", "Pediatrics" and "The Journal of pediatrics"). The unit would then have an increase of their estimated JIF with about 54% (the average increase in JIF in the three journals), also because of the high citation rate of covid-19 publications. This shows that the evaluation of the unit is affected by factors outside the unit and that the change of the normalized citation rate and JIF have no or little relation to the quality of the research at the unit or the impact the unit has at the rest of the research community.

There are some limitations of the current study. I have mainly focused on cited-side normalization of citations using the Web of Science journal categories. Other approaches for citation normalization may not be affected the same way by the covid publications, for example if citation clusters are used as fields (Colliander, 2015; Ruiz-Castillo & Waltman, 2015; Waltman & van Eck, 2013). Moreover, the field normalized citation score is an average-based indicator. Percentile-based indicators may be less affected (Bornmann et al., 2013). Furthermore, I have focused on one journal indicator only, namely the journal impact factor. I expected the JIF to be affected by the rapid growth of covid-19 publications because the indicator is based on citations from one publication year only. Other journal indicators may not be as heavily affected. Future work may focus on other indicators.

The extreme growth of covid-19 publications and their unusual citation pattern has caused citation indicators to be more unstable. There is a considerable risk to draw misleading conclusions from citation indicators spanning over the beginning of the covid-19 pandemic, in particular when time series are used and when the biomedical literature is assessed. The study supports the concern that there is a risk of overemphasizing on covid-19 related research at the expense of other research areas, not the least since citation indicators influence research policy.

**Funding information**
Peter Sjögårde was funded by The Foundation for Promotion and Development of Research at Karolinska Institutet.